\documentclass[namedreferences]{solarphysics}
\usepackage[optionalrh]{spr-sola-addons} 
\usepackage{graphicx}        
\usepackage{color}           
\usepackage{hyperref}

\newcommand{\BE}{\begin{equation}}
\newcommand{\EE}{\end{equation}}
\newcommand{\BA}{\begin{eqnarray}}
\newcommand{\EA}{\end{eqnarray}}
 \newcommand{\fig}[1]{Figure~\ref{fig:#1}}

 \newcommand{\sect}[1]{Section~\ref{sect:#1}}
 
 \newcommand{\eq}[1]{Equation~(\ref{eq:#1})}

\newcommand{\rmd}{ {\mathrm d} }
\newcommand{\uvec}[1]{ \hat{\bf #1} }
\renewcommand{\vec}[1]{\bf #1}

\newcommand{\eg}{\textit{e.g.}}

\newcommand{\ie}{\textit{i.e.}}
\newcommand{\insitu}{{\it in-situ}}
\newcommand{\todash}{\,--\,}

\newcommand{\degree}{^\circ} 
\newcommand{\dNsdR}{{\mathrm d} N/ {\mathrm d} R}

\newcommand{\dNsdRobsSD}{{{ {\mathrm d} N_{\rm obs}/ {\mathrm d} R }}}

\newcommand{\dNsdRobsFSD}{{ \frac{ {\mathrm d} N_{\rm obs} }{ {\mathrm d} R}}}
\newcommand{\dNsdRtotSD}{{ {\mathrm d} N_{\rm total}/ {\mathrm d} R }}
\newcommand{\dNsdRtotFSD}{{ \frac{ {\mathrm d} N_{\rm total} }{ {\mathrm d} R}}}
\newcommand{\dNsdRtotMCSD}{{ {\mathrm d} N_{\rm total, MC}/ {\mathrm d} R }}
  
\newcommand{\lA}{\lambda}
\newcommand{\Nbin}{N_{\rm bin}}
\newcommand{\Nfr}{N_{\rm FR}}
\newcommand{\Npart}{N_{\rm part}}
\newcommand{\Ntot}{{ N_{\rm total} }}
\newcommand{\pA}{\phi}
\newcommand{\tA}{\theta}

\begin{document}

\begin{article}

\begin{opening}

\title{Are There Different Populations of Flux Ropes in the Solar Wind?}

\author{M.~\surname{Janvier}$^{1,2}$\sep
        P.~\surname{D\'emoulin}$^{1}$\sep
        S.~\surname{Dasso}$^{3,4}$      
       }
\runningauthor{M. Janvier et al.}
\runningtitle{Flux Rope Populations in the Solar Wind}

   \institute{
 $^{1}$ Department of Mathematics, University of Dundee, Dundee DD1 4HN, Scotland, United Kingdom
\\
        email: \url{mjanvier@maths.dundee.ac.uk} \\ 
 $^{2}$ Observatoire de Paris, LESIA, UMR 8109 (CNRS), F-92195 Meudon Principal Cedex, France \\
 $^{3}$ Instituto de Astronom\'\i a y F\'\i sica del Espacio, UBA-CONICET, CC. 67, Suc. 28, 1428 Buenos Aires, Argentina \\ 
 $^{4}$ Departamento de F\'\i sica, Facultad de Ciencias Exactas y Naturales, 
UBA, Buenos Aires, Argentina \\
              }

\begin{abstract}
Flux ropes are twisted magnetic structures, which can be detected by \insitu\ measurements in the solar wind. 
However, different properties of detected flux ropes suggest different types of flux-rope population. 
As such, are there different populations of flux ropes? The answer is positive, and is the result of the analysis of four lists of flux ropes, including magnetic clouds (MCs), observed at 1~AU. 
The \insitu\ data for the four lists have been fitted with the same cylindrical force-free model, which provides an estimation of the local flux-rope parameters such as its radius and orientation.
Since the flux-rope distributions have a large dynamic range, we go beyond a simple histogram analysis by developing a partition technique that uniformly distributes the statistical fluctuations over the radius range. By doing so, we find that small flux ropes with radius $R<0.1$~AU have a steep power-law distribution in contrast to the larger flux ropes 
(identified as MCs), which have a Gaussian-like distribution.
Next, from four CME catalogs, we estimate the expected flux-rope frequency per year at 1~AU.
We find that the predicted numbers are similar to the frequencies of MCs observed \insitu. 
However, we also find that small flux ropes are at least ten times too abundant to correspond to CMEs, even to narrow ones. Investigating the different possible scenarios for the origin of those small flux ropes, we conclude that these twisted structures can be formed by blowout jets in the low corona or in coronal streamers.

\end{abstract}
\keywords{Coronal Mass Ejections; Coronal Mass Ejections, Interplanetary; Magnetic fields, Interplanetary; Solar Wind}
\end{opening}


\section{Introduction}
     \label{sect:Introduction} 

The data from \insitu\ measurements made by spacecraft in the heliosphere can show a coherent rotation of the magnetic field over time scales of hours to days. These coherent magnetic structures are typically interpreted as twisted magnetic configurations, commonly called flux ropes (hereafter FRs). 

The earliest discovered and most commonly identified FRs were the large ones, with sizes around $0.1$ AU at Earth's orbit (\opencite{Lepping90} and references therein). These FRs are called magnetic clouds (MCs) and they are characterized by a smooth rotation of an enhanced magnetic field and a proton temperature lower than the averaged solar wind while traveling with the same speed (see, \eg, \opencite{Elliott05}; \opencite{Demoulin09d}).  Their physical properties have been analyzed in depth, (see, \eg, \opencite{Dasso05}; \opencite{Lynch05}; \opencite{Lepping10}). They are seen in a fraction ($\approx$1/3) of the interplanetary coronal mass ejections (ICMEs) that are defined from a broader set of criteria (including proton temperature, ionisation levels, composition; see \opencite{Richardson10}, and references therein).   With their coronagraphs, heliospheric imagers and \insitu\ measurements, the twin spacecraft of the {\it Solar Terrestrial Relations Observatory} (STEREO) have unambiguously associated ICMEs with CMEs launched from the Sun (\opencite{Rouillard11b}, and references therein).

 Similarly, small events are also detected with heliospheric imagers \cite{Rouillard10,Sheeley10} as well as in the corona, including eruptions from ephemeral regions and narrow CMEs \cite{Mandrini05,Robbrecht09b,Schrijver10}, blowout X-ray jets (\citeauthor{Moore10}, \citeyear{Moore10},  \citeyear{Moore13}), and streamer blobs \cite{Wang09,Sheeley09}. 
In parallel, much smaller FRs than typical MCs have been detected with \insitu\ measurements \cite{Moldwin00}. Contrary to MCs, these structures present no significant variations of the proton temperature when being crossed by spacecraft. In the following, we simply call them ``small FRs'' while we use ``FRs'' to refer to all flux ropes detected \insitu. The small FRs are fitted with the same flux rope model as for MCs, by simply rescaling them in size \cite{Feng07,Cartwright08}. Statistical studies of small FRs at various solar distances indicate that small FRs expand as a power law of the solar distance, but with an exponent that is about half the exponent found for MCs \cite{Cartwright10,Gulisano10}.  
However, contrary to MCs, this expansion is not detected with \insitu\ observations of small FRs
\cite{Moldwin00,Feng07,Cartwright08,Kilpua12}.

Then, a question that arises from these different properties is whether MCs and small FRs have the same origin.
\inlinecite{Moldwin00} claimed that there are no intermediate sizes between MCs and small FRs. Furthermore, \citeauthor{Cartwright08} (\citeyear{Cartwright08}, \citeyear{Cartwright10}) argued that the FR distribution is double-peaked with different characteristics for MCs and small FRs from the \insitu\ data (such as the proton temperature and the expansion rate).  
On the contrary, \inlinecite{Feng07} found a continuous distribution in FR diameters, and \inlinecite{Wu08} found a power law for the total-energy spectrum with a slope comparable, while smaller, to the slope found in the energy spectrum of solar flares. This debate is also present at a deeper level for the definition of small FRs \cite{Feng10b,Cartwright10b}: indeed, the lists of small FRs of \inlinecite{Feng07} and \inlinecite{Cartwright08} have few cases in common, while they have a large temporal overlap and are observed by the same spacecraft (Wind).  
  
In the present study, we propose a deeper analysis of the spectrum of FRs present in the SW from different readily available lists. In \sect{Data}, we describe the four data sets used and their modeling. Then, in \sect{Populations}, we analyze the distribution of the FR radius, introducing a new technique for the statistical analysis and in \sect{localtoglobal} we analyze and correct the biases of the observed distributions. This technique is applied in \sect{SFR} to study the small FRs. 
In \sect{MC}, we separate MCs from small FRs, then in \sect{related} we provide evidence that only MCs are related to CMEs. Finally, we investigate the possible origin of small FRs in \sect{origin_SFR}. 
We summarize our results and conclude in \sect{Conclusion}.

\section{Description of the Four Flux Rope Lists} 
      \label{sect:Data}      
   In the present study we use four lists of FRs observed in the solar wind nearby Earth by the Wind or {\it Advanced Composition Explorer} (ACE) spacecraft.
  The first list of FRs, that are mostly MCs, is attached to the article of \inlinecite{Lynch05}.
It contains $132$ FRs observed from February 1995 to November 2003.  
   A second list of FRs is provided in Table~1 of \inlinecite{Feng07}.
It contains $144$ FRs (of both MCs and small FRs) observed from February 1995 to September 2001.
   A third list of small FRs is given in Table~1 of \inlinecite{Feng08}.
It contains $125$ small FRs observed from February 1995 to November 2005.
   Finally a fourth list of MCs (Table~2 at \href{http://wind.nasa.gov/mfi/mag\_cloud\_S1.html}{wind.nasa.gov/mfi/mag\_cloud\_S1.html}) is based on the results of \inlinecite{Lepping10} and it provides an extension of the list to more recent MCs.  
This list, as of 12 July 2013, contains the parameters obtained for $121$ MCs observed 
by the Wind spacecraft from February 1995 to December 2009.
The four lists have a comparable number of FRs, 
but every list spans different spatial and temporal intervals (although they partly overlap). In the following, we study them independently.   
 
   Since MCs typically have a low plasma $\beta$, they are considered to be in a near force-free-field state. Moreover, since clear FR signatures are present in the \insitu\ data, their magnetic configuration is generally modeled using force-free FR models. The simplest, but still widely used model, is the cylindrical linear force-free field \cite{Lundquist50}:
   
  \begin{equation}  \label{eq:Lundquist}
  \vec{B}_L = B_0 [J_1(\alpha r) \uvec{e}_{\phi} + J_0(\alpha r) \uvec{e}_{z}] \,,
  \end{equation}
where $J_0$ and $J_1$ are the ordinary Bessel functions of order $0$ and $1$, and $\uvec{e}_{\phi}$ and $\uvec{e}_{z}$ are the unit vectors in cylindrical coordinates.
The axial component of $\vec{B}_L$ is typically imposed to vanish at the boundary $r=R$ of the FR, so $\alpha$ is determined by the first zero of $J_0$, implying $\alpha \approx 2.4/R$.

   The Lundquist model is then used to fit each event in the four lists described above.
The least-square-fitting procedure is the same as, or closely similar, to that used by \inlinecite{Lepping90}.
Variations of the fitting procedure include the precise algorithm to achieve the non-linear fit and the choice of the FR boundaries.  We refer to the articles cited above for further information on the implementation of the fitting procedure for each list of FRs.
The fit to the \insitu\ data determines the seven free parameters of the model (the longitude and the latitude of the FR axis, the distance of the spacecraft from the FR axis at closest approach point, the magnetic-field strength on the FR axis, the twist, the sign of the magnetic helicity, and the time at closest approach to the FR axis).
The ones used in the present study are: the flux-rope axis orientation (longitude [$\pA$] and  latitude [$\tA$]), the magnetic field strength on the FR axis [$B_0$], and the FR radius [$R$] deduced from the previous parameters.

\section{A New Technique for the Analysis of the FR Distributions} 
      \label{sect:Populations}      

  \inlinecite{Cartwright08} found evidence that the distribution of FRs is double-peaked with the most numerous FRs having durations below four hours while the second peak of the distribution, around 12\todash 16 hours, corresponds to MCs. In the following, we further analyze the FR distributions by using the four different lists described above. The temporal duration of a flux-rope crossing depends both on its speed and on its orientation (with a longer crossing time as the spacecraft crossing trajectory is nearly aligned with the FR axis). On the other hand, the FR radius is an intrinsic property of the FR. Therefore, we study in the following only the distributions of the FR radius.

A classical statistical analysis of the distribution of a parameter (here the radius [$R$]) involves histograms with a uniform binning width. 
The key point of such an analysis is to define a proper number of bins [$\Nbin$]. 
Indeed, the bin width needs to be not too narrow, in order to have enough counts in each bin, but needs to not be too broad either, in order to represent as well as possible the distribution variations.
However, when this distribution has a broad range of counts in various bins over a large range of $R$, the most suitable bin width cannot be found. This is the case for observed FRs as the selection of a bin width adapted to the MCs implies that small FRs are all set in a few bins (see, \eg, Figure 10 of \opencite{Cartwright08}).

In order to analyze a distribution extending over a large range of radius with large count variations, the bin size should be adapted either to the radius scale or to the FR counts. 
We select the last option as it provides a uniform count, so it decreases and spreads uniformly the statistical fluctuations over all the bins.
We proceed by first ordering the FRs by growing radius.  Next, we define partitions with $\Npart$ cases in each.  The lowest partition contains the $\Npart$ cases with the lowest radius. In the next one, the cases are shifted to higher radius by one case only, and so on until the case with the largest radius is incorporated in the last partition. We choose to shift from one partitition to the neighboring one by only one case, as the method can be applied to any total number of cases $\Nfr$ without removing any data (\eg\ a shift with two cases would remove the FR with the largest radius if $\Nfr$ is odd).  
Contrary to a classical histogram, where a FR is only present in one bin, the partition method implies that FRs become present in consecutive partitions apart from the first and last ranked FRs.
This implies a larger sampling of the distribution functioccomn allowing to better identify its variations (because the number of partitions can be much larger than the number of bins of an histogram).

The distribution $\mathcal{D}(R)$ deduced from a histogram is given by the number of cases [$\Delta N=N(R_2)-N(R_1)$] in each bin of fixed width [$\Delta R=R_2-R_1$]:
  \begin{equation}  \label{eq:distribution}
  N(R_2)-N(R_1) = \int_{R_1}^{R_2} \mathcal{D}(R) \textrm{d}R 
            \approx \mathcal{D} \bigg(\frac{R_1+R_2}{2} \bigg) (R_2-R_1) \,.
    \end{equation}

With the partition technique presented above, $\Delta N$ is fixed to a value $\Npart$ while $\Delta R$ remains variable. $\Delta R$ can be estimated by the boundary values of $R$ for each partition.
Performing several tests with randomly generated distributions, we verified that a robust estimation of $\Delta R$ is to define it with the standard deviation of $R$ [$\sigma_R$], for each partition. 
More precisely, the two quantities are linked by: $\Delta R = 2\sqrt{3} ~\sigma_R $ supposing a uniform distribution in each partition. 
Using randomly generated distributions confirms that the value of $\sigma_R$ in a partition is weakly affected by the type of analyzed distribution provided that the partition extension is small compared with the variation of the global distribution.  
This was also tested with a broad range of synthetic power-law distributions (see Section \ref{sect:SFR}), by computing analytically $\sigma_R$ and comparing it to the numerical results.   We conclude that the estimation of $\Delta R$ with $\sigma_R$ is reliable enough to provide an estimation of the distribution in each partition. In the following, we analyze the distribution [$\dNsdR$] estimated with $\Npart / (2\sqrt{3} ~\sigma_R)$ for each partition, versus the mean value $\left< R\right>$, hereafter simply denoted $R$ in the graphs for convenience.

\begin{figure}  
 \centerline{ \includegraphics[width=0.8\textwidth]{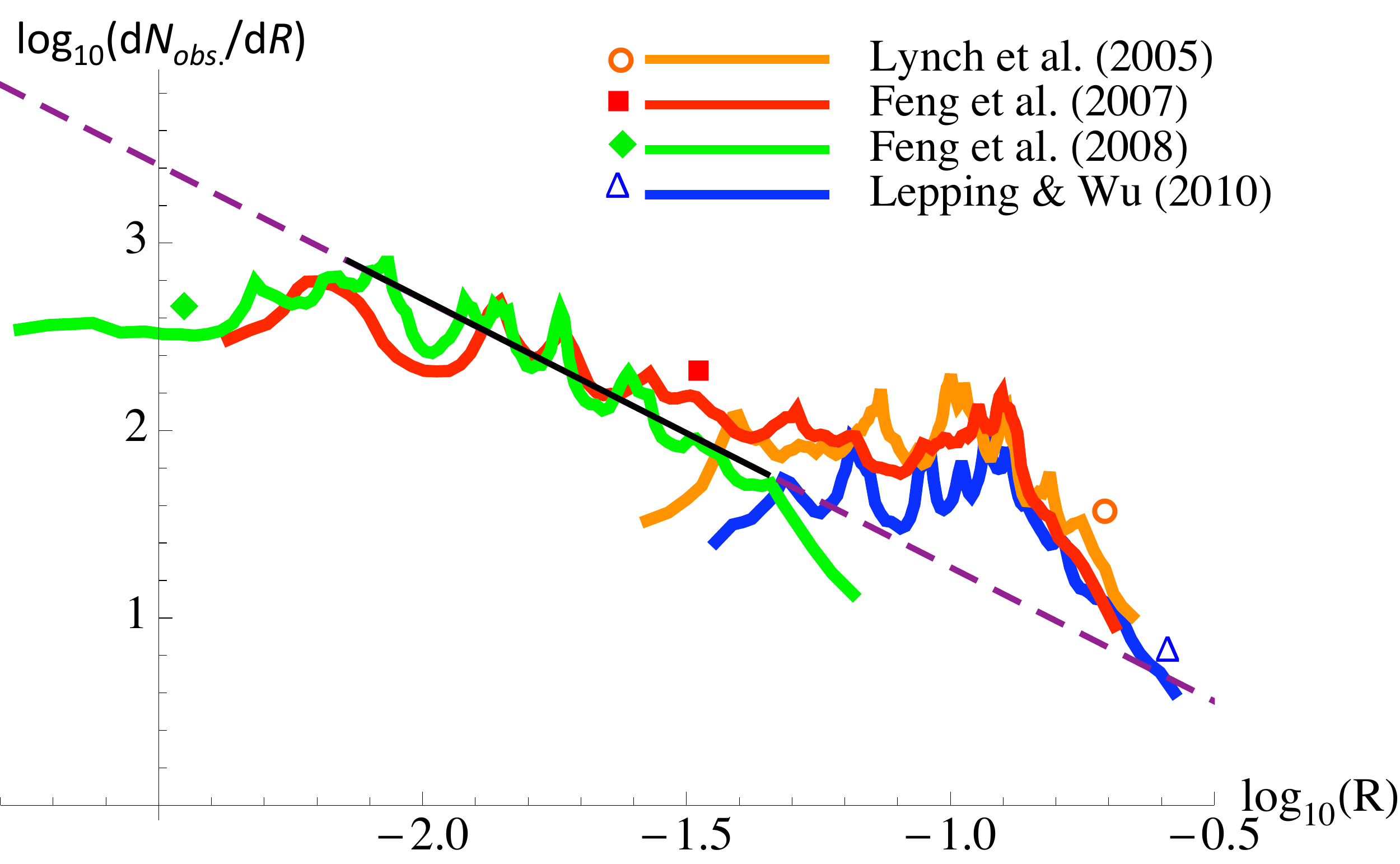} }
\caption{Observed [$\dNsdRobsSD$] distributions of flux ropes versus the mean FR radius of each partition (see Section \ref{sect:Populations} for the partition technique), noted here $R$ for simplication, in log\todash log scales.  $N$ is the number of cases per year, with a number of cases in each partition being $\Npart = 10$, and the FR radius [$R$] is in AU. The continuous black straight line is the fit of the distribution of small FRs from Feng et al. (2008) within the radius range $[0.008,0.05]$~AU. The fit is extended outside the fitting region by the dashed purple line.
}
 \label{fig:dNsdR}
\end{figure}

As the four lists are associated with different time ranges (\sect{Data}), we normalized the obtained FR distribution with the number of related years of observation, to be able to compare all the results.  
Their distributions are shown in \fig{dNsdR}, for which we have chosen $\Npart=10$ so as to minimize the averaging effect (\ie\ its possible bias consequences).  This low $\Npart$ value implies statistical fluctuations of $\Delta R$ of amplitude $\lesssim 0.2$ in the logarithm.  Nevertheless, looking at the distributions of FRs from the different lists, there is a large bump around $R=0.1$~AU that stands well above the power-law distribution (indicated in black) present for smaller FRs and that cannot be explained simply by statistical fluctuations. The three lists which include MCs nearly agree for large $R$ values (red, orange, and blue curves in \fig{dNsdR}), although the list of \inlinecite{Lepping10} (blue curve) excludes a significant number of MCs below $R\approx0.1$~AU contrary to the lists of \inlinecite{Lynch05} and \inlinecite{Feng07}.
Finally, a power-law fit of the list of \inlinecite{Feng08} (black line) that is limited to small FRs, shows that the larger FRs, \ie\ the MCs, are much more numerous than expected with the extension of this power law to larger radius.  We conclude that there are two populations of FRs with different origins (as their distributions are clearly different).   We further analyze these two distributions in the next two sections.

\section{From Local to Global Probability Distributions} 
      \label{sect:localtoglobal}

In the intermediate range of radius, $\dNsdRobsSD$ is close to a power law as fitted by the straight line in the log\todash log plot of \fig{dNsdR}.  However, this power law seems to break down for small radius, especially for the list of \inlinecite{Feng08} which contains the largest number of small FRs.
Then, we investigate whether the global parameters of such small FR have any peculiarities and whether selection effects play a role in the breaking down of the power law. To do so, we first ordered the FRs with one of the global parameters, and then we split the FRs in few groups, with growing values of the selected parameter. This allows us to test whether the distributions $\dNsdRobsSD$ is affected by this selected parameter. As we found a significant effect only with the location angle [$\lambda$], we describe the results obtained below.

\begin{figure}  
 \centerline{ \includegraphics[width=1.\textwidth]{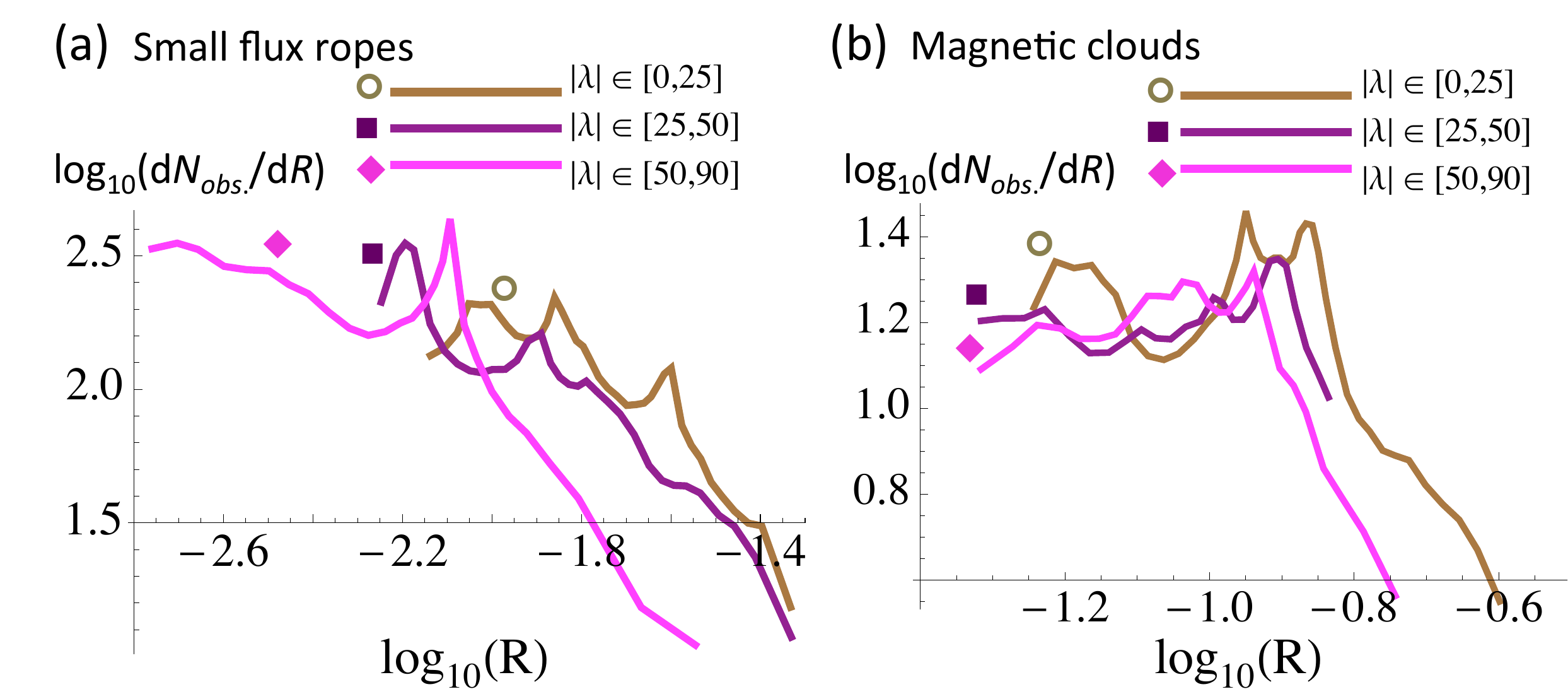} }
\caption{Observed FR distributions [$\dNsdRobsSD$] versus FR radius in log\todash log scales split in groups of $|\lA|$ (angle between the flux rope axis and the plane orthogonal to the Sun--spacecraft direction, see \eq{lA}). The ranges of $|\lA|$ are selected to have a similar number of FRs for each distribution.
{\bf (a)} Distributions for small FRs (Feng et al. 2008).
{\bf (b)} Distributions for MCs (Lepping and Wu, 2010).
}
 \label{fig:lambda}
\end{figure}

	The location angle [$\lA$] is the angle between the FR axis direction and the direction orthogonal to the Sun \todash spacecraft direction (which is the typical direction of the FR propagation; see the schema in Figure 1 of \opencite{Janvier13}). This angle is referred to as a location angle as it directly gives an idea on the position of the spacecraft along the FR axis.
Note that $\lA$ is related to the longitude [$\pA$] and latitude [$\tA$] of the FR axis by the following relation:     
  \BE
  \sin \lA = \cos \pA ~\cos \tA    \label{eq:lA} \,.
  \EE
The location angle [$\lA$] becomes 0 for an orthogonal crossing of the FR (spacecraft at the apex of the FR), and $|\lA|$ grows as the crossing is more along the FR axis. 
Then, we split the data of \inlinecite{Feng08} into three groups according to their $|\lA|$ values. 
The number of groups is limited by the total number of FRs, and we choose the boundaries of the three groups so as to approximately have the same number of FRs and for the corresponding $\dNsdRobsSD$ to be comparable in magnitude and in statistical fluctuations. The amplitude of the statistical fluctuations seen throughout the three distributions shows that the central bump in the distribution of $50^{\circ} < |\lA| \le 90^{\circ}$ (in purple in \fig{lambda}a) is likely to be a statistical fluctuation. \fig{lambda}a shows that the slope for large $\log R$ is comparable for the three different groups, while the group with the largest $|\lA|$ values have much more numerous smaller FRs.    We interpret this result as a selection effect on the duration of observed FRs: for the same duration and velocity, the FRs more inclined along the Sun--spacecraft direction (\ie\ with larger $|\lA|$ values, as the spacecraft crossing is more along the FR axis) have a smaller radius.   We conclude that the breakdown of the power law for the smaller radius, as shown in \fig{dNsdR} with the tail of the green curve from \inlinecite{Feng08}, is due to a selection effect. As such, we exclude from the power-law fit the FRs with $R<0.008$~AU.  Such a selection effect is not present in the MC list of \inlinecite{Lepping10} as shown in \fig{lambda}b where similar distributions for different groups of MCs ordered by their related $\lA$ are shown.

Other selection effects can play a role in the distribution of the flux ropes. Indeed, the probability of FR detection is proportional to the apparent surface of the FR, which is defined as the surface area of the FR projected on the sphere centered on the Sun and crossing the spacecraft location (the FR motion being almost along the Sun--spacecraft direction).
The smaller a FR is, the smaller its apparent surface is, so the smaller its probability of being crossed by the observing spacecraft is.  The studied FRs are all observed at a distance $D=1$~AU.  The surface of a flux rope projected on the sphere $r=D$ is $\approx 2RL_{\rm p}$, where $L_{\rm p}$ is the length of the axis projection (on $r=D$) and where we assume a similar radius $R$ along the FR axis when observed at 1~AU (as expected from the axial flux conservation along the flux rope and an approximate total-pressure balance with the surrounding solar wind).  This implies that small FRs are under-represented in the FR lists at least by a factor proportional to $1/R$ \cite{Wu08}, and probably more as $L_{\rm p}$ is expected to be an increasing function of $R$ (see below).  

   Next, we estimate the probability of detecting a FR.  We suppose that the FRs have an equiprobability of presence in the whole range of ecliptic longitude and in a band of latitude $\pm \theta_{\rm max}$.   Indeed, the numerous rotations of the Sun and its associated interplanetary magnetic field, on time scale of several years, imply that FRs formed either in the corona or in the heliospheric current sheet can be detected at any longitude. However, the equiprobability hypothesis in latitude is more challenging in both formation scenarios, and as such is only the simplest hypothesis.   Within this framework, the probability of detecting one FR [$P_{\rm FR}$] is the ratio of the solid angles, as seen from the Sun, of the FR and of the latitude band considered, as follows:
   \BE \label{eq:proba}
   P_{\rm FR} = \frac{2 R L_{\rm p} }{4 \pi \sin \theta_{\rm max} D^2} \,.
   \EE

Then, the distribution $\dNsdRobsSD$ for detected flux ropes needs to be corrected by a factor $1/P_{\rm FR}$ in order to estimate the total number of FRs released per year.   For small FRs, we have presently no information on $L_{\rm p}$, while for MCs the heliospheric imagers indicate that $L_{\rm p}$ is comparable to $D$ (see \eg\ \opencite{Janvier13} and references therein).  We then take $L_{\rm p}=D$ in the following numerical applications.  We anticipate that this value is too large for the small FRs as these thin structures are not expected to keep their twisted structure over such large distances, as they can be affected by reconnection with the surrounding solar wind magnetic field, as found in some MCs \cite{Dasso06,Dasso07,Ruffenach12}.  In particular, $L_{\rm p}$ is expected to decrease for FRs with smaller $R$ as these flux ropes are more likely to be affected by reconnection as they have a lower axial magnetic flux. For example, if $L_{\rm p}$ behaves as a power law of $R$, such as $R^{s_{\rm L}}$, the corrected distribution function [$\dNsdRtotSD$] would have to be multiplied by $R^{-s_{\rm L}}$, so that its slope (on a logarithmic scale) would decrease by $-s_{\rm L}$. Instead, in the following, we implement a conservative correction by correcting $\dNsdRobsSD$ only with $R$ (so we keep $L_{\rm p}=D=1$ AU for all FRs). 
Then, the total distribution [$\dNsdRtotSD$] of the FRs present at 1~AU per year is a function of $R$ as
   \BE \label{eq:dNsdRtot}
   \dNsdRtotFSD = \dNsdRobsFSD  \frac{2 \pi \sin \theta_{\rm max}}{R/D} \,.
   \EE

The range of latitudes $\pm \theta_{\rm max}$ where FRs are present cannot be derived by \insitu\ data but only from coronagraphs and heliospheric imagers.  So far, this is only possible for MCs which have been associated to CMEs (\opencite{Rouillard11b}, and references therein). 
The CME latitude range is solar-cycle dependent and to a lesser extent dependent on the CME catalog used \cite{Yashiro04,Robbrecht09b}.  Moreover, the measured CME latitudes are subject to large projection effects (when they are not ejected near the limb) and the data are providing only an upper limit on the true latitude of a CME.  Locating the source regions of CMEs allows us to better define the latitude distribution of CMEs.  The CME source regions are typically found in the active-region belt, which extends up to $40\degree$--$50\degree$ of latitude in both hemispheres with a shift to lower latitude during the solar cycle (see \opencite{Cremades04}, Figure 9; \opencite{Tripathi04}, Figures 11, 16, and 18; \opencite{Howard08}, Figure 2c; \opencite{Wang11}, Figures 3 and 4a). However, CMEs are not all moving out radially and moreover some are deflected by nearby coronal holes, especially towards the Equator during solar minimum \cite{Cremades04,Cremades06,Wang11}.  It implies that the distribution of source regions is only an approximation of the distribution of the CME propagation directions. 
In the following results, we use $\theta_{\rm max}=45 \degree$ as an estimation of the mean extension of the CME propagation direction.
Taking into account these limitations, the quoted numbers would simply be multiplied by a factor 0.8 (resp. 1.4) if $\theta_{\rm max}=30 \degree$ (resp. $60 \degree$) would be used instead.

\begin{figure}  
 \centerline{ \includegraphics[width=1.\textwidth]{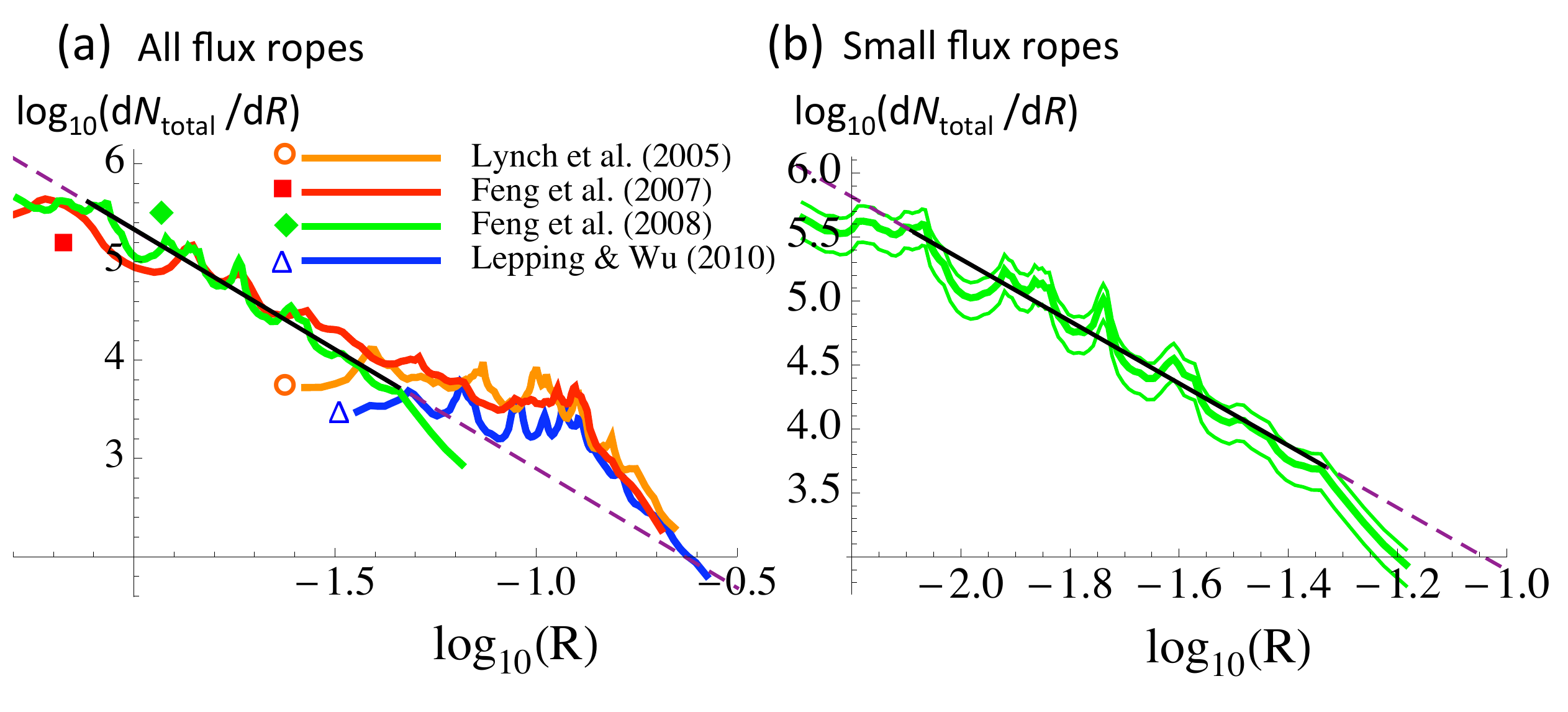} }
\caption{FR distributions, corrected from the FR surface selection effect, versus the FR radius in log--log scales. 
(a) The continuous black straight line is the fit of the distribution of small FRs from Feng et al. (2008) within the radius range $[0.008,0.05]$~AU. The fit is extended outside the fitting region by the dashed purple line.
(b) Corrected distributions for small FRs (Feng et al., 2008). The central curve, also shown in (a), is surrounded by the distributions computed with $\Npart \pm \sqrt{\Npart}$, where $\Npart=10$ is the number of FRs in each partitions, in order to show the effect of statistical fluctuations in each partition.
}
 \label{fig:dNsdRcor}
\end{figure}  

\section{Corrected Population of Small Flux Ropes} 
      \label{sect:SFR}      

The distributions corrected from the biases described in Section \ref{sect:localtoglobal} are shown in \fig{dNsdRcor}, and they show that small FRs are much more numerous than large ones, especially MCs.  The power law is steep with a slope of $-2.4 \pm 0.14$ where the slope error is computed with a confidence level of 95\%. This means that for a decrease of $R$ by a factor of ten, there are $\approx 250$ times more FRs.  Moreover, in the present data, there is no indication that the power law would stop at the lowest radius shown ($0.005$~AU) 
since the flattening of the distribution for low $R$-values was identified as a bias due to a selection of the axis orientation (\fig{lambda}).   There is also no indication that the power law stops, or that its slope changes at larger $R$ values before being covered by the MC distribution. 
   
A deeper look at \fig{dNsdRcor}a shows a possible temporal evolution of the power law slope: a smaller slope is computed from the list of \inlinecite{Feng07} (red curve in \fig{dNsdRcor}a) compared to the one computed from the more extended list of \inlinecite{Feng08} (green curve in \fig{dNsdRcor}a).  To investigate this temporal effect more thoroughly, we split the list of \inlinecite{Feng08} into three temporal intervals with similar number of small FRs.   Keeping the same range of $\log_{10} R$ for the fit, we find slopes of $-1.8 \pm 0.1$, $-2.1 \pm 0.5$, and $-3.4 \pm 0.3$ for the time ranges [1995,1997],
[1998,2001], and [2002,2005], respectively.  This indicates a plausible temporal evolution of the power-law slope. However, a much larger number of small FRs studied over a much longer period of time would be needed to confirm this and to test if this evolution is related to the solar cycle.  

\begin{figure}  
 \centerline{ \includegraphics[width=1.\textwidth]{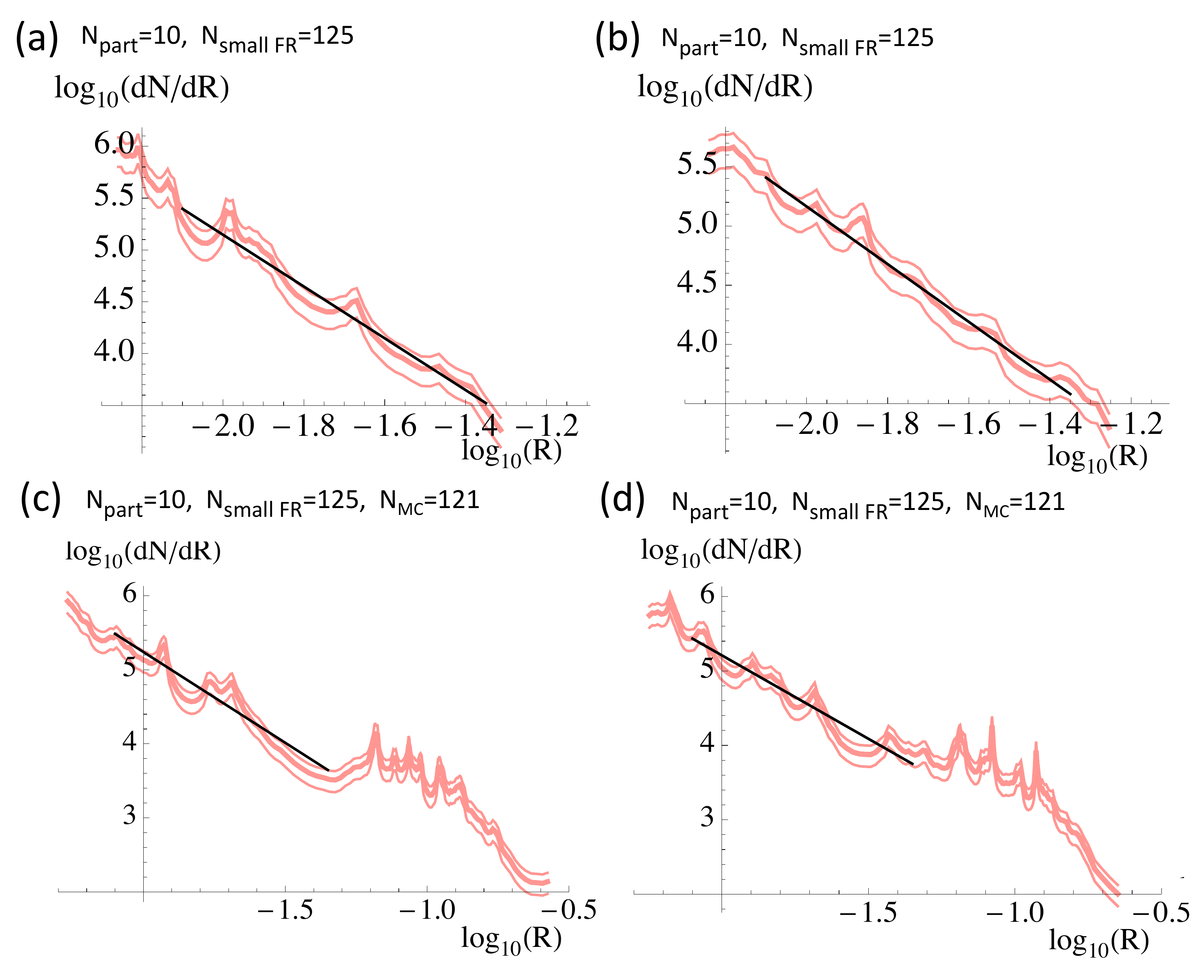} }
\caption{(a,b) Tests with a power-law distribution $R^{-2.4}$ analyzed similarly as the observations (\fig{dNsdRcor}). Panels a and b are two different realizations with the same number of cases and partitions as in the observations.  The central curves are surrounded by the distributions computed by $\Npart \pm \sqrt{\Npart}$  to show the effect of statistical fluctuations in each partition. 
(c,d) Other realizations with a Gaussian distribution in $\log _{10} (R)$ added to simulate MCs (mean value at $\log _{10} (R)$ = -1 and standard deviation $= 0.16$). They show similar amplitude of fluctuations than the observed distributions shown in \fig{dNsdRcor}.
}
 \label{fig:dNsdRtest}
\end{figure}  

To complete the study of the observed distributions of flux ropes in the solar wind, we tested the partition technique used so far to derive $\dNsdRtotSD$ with a power-law distribution [$R^{-s}$] generated by a random generator. We set the range of $R$ to the observed range of flux ropes radius from the list of \inlinecite{Feng08} (\fig{dNsdRcor}b) and $s=-2.4$ so that the tests are directly comparable to the observations.  
For a larger number of cases in the synthetic distribution, the slope of the fit becomes closer to $s=-2.4$ and the estimated error on the fitted slope decreases accordingly, as expected.  

The most relevant test consists of a number of cases similar to the number of observed FRs. Then, we select $N=125$ to compare with the results obtained with the list of \inlinecite{Feng08}.  The tests are also shown with the same number of cases per partition [$\Npart$] as for the observations.  The results of two typical tests of the theoretical distributions are shown in \fig{dNsdRtest}a,b. From the fitting of the curves, the slope value is generally found to be around $-2.4$ with variation from case to case by less than $0.2$.  The test results also show statistical fluctuations of $\dNsdR$. They are comparable in magnitude to the ones present in \fig{dNsdRcor}b, showing that such fluctuations, such as the bump in the middle of the distributions, have no physical meaning.   They can be reduced by increasing the $\Npart$ value without any significant effect on the derived slope (within the error bar quoted above).
We also performed other tests changing $s$, $\Npart$, the range of $R$ studied, and the number of cases.
All of these results show that the partition technique is a robust technique to derive the $\dNsdR$ distribution, especially when it is a steep function over a broad range of $R$-values.

As a final step, we also compare the partition technique used above with the more classical histogram technique, binning both the observed and synthetic data with $\Nbin$ bins of same extension in $R$.  We found slopes comparable with the ones quoted above but with much larger uncertainties, typically larger by a factor three to eight.   These larger uncertainties with a classical histogram are due to the bins with large $R$ values, which have a small count number so large statistical fluctuations.   With a histogram, the slope is also sensitive to the location of the bin boundaries as well as their number $\Nbin$.  For example the slope changes from $-2.8$ to $-2.5$ when $\Nbin$ is changed from 5 to 20 for the data of \inlinecite{Feng08}, while the slope reaches $-1.95$ for $\Nbin=40$.   In contrast, the slope is not significantly affected by $\Npart$ with the partition technique (it changes by less than $0.1$ when $\Npart$ is changed from 5 to 40).   This conclusion was also extended to the theoretical distributions described above.   We conclude that the partition technique is the best suited to analyze $\dNsdR$, a logical conclusion since it optimizes the distribution of the small number of observed FRs in the partitions.

\section{MC-only Distributions} 
      \label{sect:MC}      

In the original distribution of flux ropes (\fig{dNsdR}), the large flux ropes, \ie\ the MCs, stand above the power-law distribution of the small FRs and within a large bump. This bump remains even after correcting the observed distribution (see \fig{dNsdRcor}a). This bump is both higher and broader in $R$ than the fluctuations due to the small statistics in the partitions (with comparable fluctuations in observations, \fig{dNsdRcor}, and in theoretical tests, \fig{dNsdRtest}).     

\begin{figure}  
 \centerline{ \includegraphics[width=1.\textwidth]{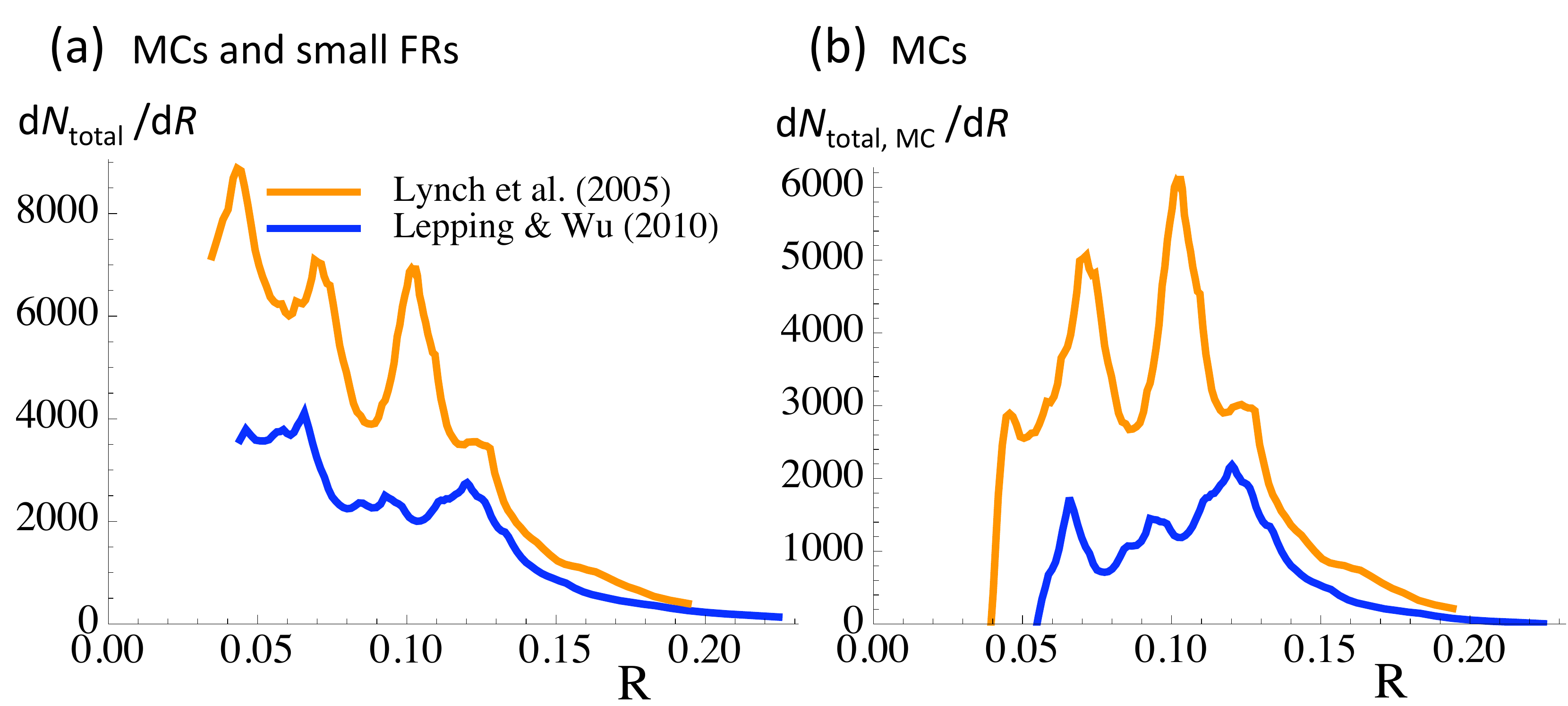} }
\caption{FR distributions corrected from the FR-surface-selection effect, versus the FR radius. We choose $\Npart =20$
to limit statistical fluctuations.  
(a) Distributions with all the FRs present in the reported lists.
(b) A power-law fit to the small FRs (shown in \fig{dNsdRcor}b) is subtracted to the distribution shown in panel a. Then panel b shows an estimation of the distribution limited to MCs only. (the different peaks are likely to come from statistical fluctuations, see Section \ref{sect:MC})}
 \label{fig:dNsdR_MCs}
\end{figure}  

The $\dNsdRtotSD$ distributions of MCs obtained from different lists are evaluated per year and therefore can be compared.  The three data sets have a similar global distribution, especially from the lists of \inlinecite{Lynch05} and \inlinecite{Feng07} which have the closest time range ([1995,2003] and [1995,2001], respectively).   The distribution of \inlinecite{Lepping10} is close to the previous ones only for $R \gtrsim 0.12$~AU (\fig{dNsdRcor}a). For lower values of $R$, the distribution becomes lower by a factor two approximately.   This can be partly explained with a temporal evolution as follows: we first limit Lepping's list to 2001 (2003) to be directly comparable to the time period of \opencite{Feng07} (\opencite{Lynch05}). 
Then, with the same time interval, $\dNsdRtotSD$ for the three MC lists is closer (within the statistical fluctuations) for $R \gtrsim 0.08$~AU.  On the other side, for $R \lesssim 0.08$~AU, the $\dNsdRtotSD$ distribution of Lepping's list is rather nearby the power law distribution of the small FRs while both other distributions are still well above (comparable to \fig{dNsdRcor}a).
This indicates a selection effect compared to the lists of \inlinecite{Lynch05} and \inlinecite{Feng07}.  These last authors have probably also included as MCs parts of the cases considered as MC-like by \inlinecite{Lepping05}. Put differently, \inlinecite{Lepping10} have set more restrictive conditions on the definition of MCs. Such disagreements among independent lists of MCs could have several origins as summarised in the introduction of Lepping et al. (2005).

Next, we extend the modeled distribution function of \sect{SFR} by adding to the synthetic power law 121 cases to represent the MCs with a Gaussian distribution in $\log R$ (\fig{dNsdRtest}c,d).  Various realizations of this distribution show fluctuations of the same order than in \fig{dNsdRcor}a: then, peaks in the MC region are also due to statistical fluctuations.

The comparison of lists, detailed above, shows an ambiguity in identifying an event as a small FR or as a MC within an intermediate range of radius (around $R=0.05$~AU).  A possible way to remove this ambiguity is to do a precise case by case analysis specifying that a MC should have a proton temperature significantly lower than the mean proton temperature found in the solar wind traveling at the same speed.  By significant, a factor of two can be used, as typically used more generally for ICMEs \cite{Richardson10}.  A second arbitrary parameter that has to be chosen is the fraction of the total number of FRs such that this condition on temperature should hold to define a MC.  The splitting between small FRs and MCs will depend on the values adopted for those parameters.  Here we rather take a statistical approach to disentangle the two distributions, as follows.   

Since there is no indication in the data that the power law found for small FRs (\fig{dNsdRcor}) is ending
(or changing slope) before the radius range of MCs, we assume that this power law extends to larger $R$ values than the fit range used to derive it.  Then, we substract this power law from the global $\dNsdRtotSD$ distribution to estimate the distribution of MCs only [$\dNsdRtotMCSD$]. Both the Lynch and Lepping lists have Gaussian-like distribution for $\dNsdRtotMCSD$ (\fig{dNsdR_MCs}b) when plotted with linear scales. This contrasts with the total FRs distributions having a large tail on the small $R$ side (\fig{dNsdR_MCs}a).   The mean values ($0.094$ and $0.11$~AU) and standard deviations ($0.026$ and $0.02$~AU) of $\dNsdRtotMCSD$ are comparable.  The main difference is the magnitude of $\dNsdRtotMCSD$ as earlier described in log--log plots (\fig{dNsdRcor}a).

Again, this statistical study aiming at providing the distribution of observed MCs gives strong evidence that small flux ropes and larger ones, \ie\ MCs, are of different origin.

\begin{figure}  
 \centerline{ \includegraphics[width=0.6\textwidth]{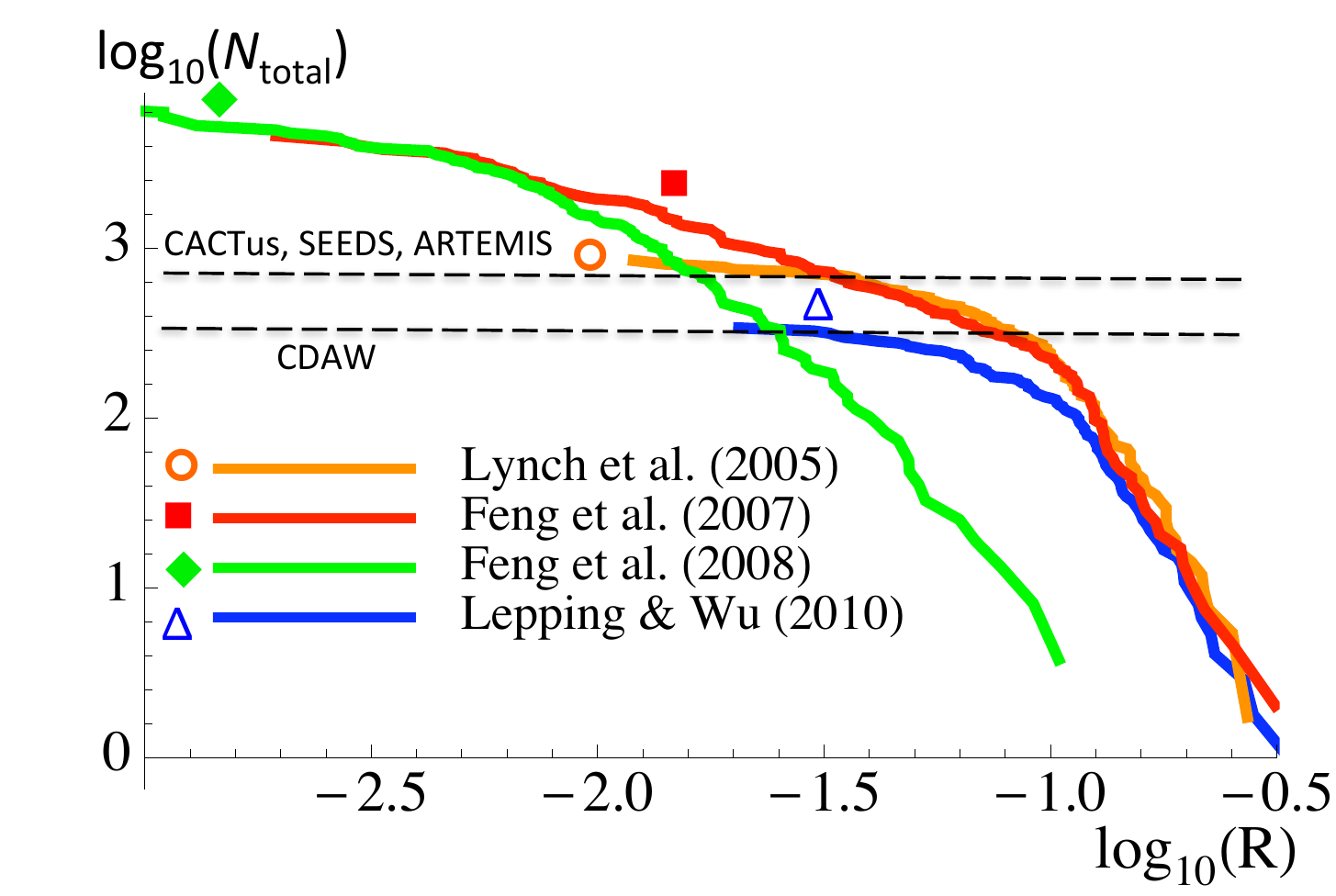} }
\caption{Cumulative distribution functions of the number of events [$\Ntot $] computed per year with an average over the time period of the lists.  The number of events in all curves is corrected from the FR apparent surface, \eq{dNsdRtot}, and the summations are done from the larger to the smaller radius, \eq{N(R)}. The number of FRs per year estimated from the CME catalogs are shown by dashed black lines (see Table~\ref{Table:MCpredicted} for values).
}
 \label{fig:dNsdRcumul}
\end{figure}  

\section{Are MCs and/or Small FRs Related to CMEs?} 
      \label{sect:related}      

We first investigate the cumulative function of cases per year [$\Ntot (R)$], doing the summation of cases from the larger to the lower radius, as
  \begin{equation}  \label{eq:N(R)}
  \Ntot (R) = \int_{R}^{R_{\rm max}} \dNsdRtotFSD ~\rmd R  \,.
  \end{equation}
   The summation of cases is done starting from the larger ones ($\Ntot (R_{\rm max})=0$) to lower radius values because the FRs with larger radius are much less numerous, \ie\ with this definition the larger FRs are not covered by the smaller FRs in the summation of cases.
   We only present the results already corrected from the apparent FR surface bias (\ie\ with $\dNsdRtotSD$).  As expected from the $\dNsdRtotSD$ curves (\fig{dNsdRcor}a), the $\Ntot (R)$ curves, derived from lists with MCs, have nearby values for the larger range of $R$ corresponding to large MCs (\fig{dNsdRcumul}).  The results of \inlinecite{Lynch05} and \inlinecite{Feng07} nearly agree down to $R \approx 0.025$~AU, while the results from \inlinecite{Lepping10} are significantly lower for small $R$ (factor $\approx 2$).  Finally, as expected, the total number of FRs is dominated by the smaller ones (Feng's lists). 

Next, we compare the above total number of FRs per year to the corresponding total number of CMEs observed during the same period of time.  
This requires estimating the number of FRs expected from the observed number of CMEs.  Recent studies conclude that a large fraction \cite{Vourlidas13} or even most of the CMEs \cite{Gopalswamy13,Makela13} have a flux rope structure.
The CMEs observed \insitu\ (ICMEs) without flux ropes would be cases crossed by the spacecraft close to the FR boundary, or even outside the FR (\eg\ \opencite{Jian06}).  The fraction of MCs found in ICMEs is solar-cycle dependent and we used the yearly fraction found by \inlinecite{Richardson10} to convert the observed number of CMEs to the predicted number of FRs.  

The early number of CMEs is counted by at least four catalogs.  The first catalog is the    
Coordinated Data Analysis Workshop (CDAW) and it relies on visual analyses of LASCO/SoHO images by operators \cite{Yashiro04}.   The next three catalogs are based on different computer algorithms to track the brightness evolution in coronagraph images.   They are the Computer Aided CME Tracking (CACTus: \opencite{Robbrecht09b}), the Solar Eruptive Event Detection System (SEEDS: \opencite{Olmedo08}),
and the Automatic Recognition of Transient Events and Marseille Inventory from Synoptic maps (ARTEMIS: \opencite{Boursier09}).   Each catalog has its own bias as described in the above citations and by \inlinecite{Yashiro08}.  Typically, the algorithm-based catalogs detect more and fainter CMEs, and some are considered as incorrect detection by CME experts. 

From the above catalogs we compute an estimation of the number of FRs, as follows.
Since the LASCO coronagraph was not working in 1995, there is no CME data.
However, the CME rate correlates closely with the sunspot number \cite{Robbrecht09b} which is at a comparable mean level in 1995 and 1997, so we duplicate the data of 1997 for 1995 (in any case, the expected number of CMEs in 1995 is   small compared to the years after 1998, so this represents a small correction).  For the years 2008 and 2009, \inlinecite{Richardson10} found no MCs, so there is no need of CME counts for these two years to predict the FR number.
From these data, the predicted number of MCs is computed per year and summed up for the time interval of the FR lists (Table~\ref{Table:MCpredicted}, columns 2--5).  The results with the algorithm-based catalogs are very close to each other as they detect a similar number of CMEs \cite{Boursier09}. 

\setlength{\tabcolsep}{5pt}   
\begin{table}   
\caption{Predicted total number (rounded values) of FRs ejected from the Sun per year during four time periods (corresponding to the FR lists, see the sixth column).
The estimations are based on four CME catalogs: CDAW, CACTus, SEEDS, and ARTEMIS (see Section \ref{sect:related}). 
The last column presents the maximum values of the cumulative function of $\Ntot $ shown in \fig{dNsdRcumul}, averaged per year over the time period indicated, \ie\ the estimated total number of FRs present per year at 1 AU.
}
\label{Table:MCpredicted}
\begin{tabular}{ccccccc}     
  \hline                   
time period & CDAW & CACTus & SEEDS & ARTEMIS & reference of FR list & $\Ntot$ \\
  \hline
1995--2003 & 330 & 790 & 740 & 720 & Lynch \textit{et al.} (2005)   & ~\,850 \\
1995--2001 & 290 & 680 & 610 & 630 & Feng \textit{et al.} (2007) & 4600 \\
1995--2009 & 370 & 680 & 580 & 510 & Lepping \& Wu (2010) &  ~\,390\\
1995--2005 & 330 & 760 & 710 & 660 & Feng \textit{et al.} (2008) &  7040 \\
  \hline       
\end{tabular}
\end{table}   

Next, we compare the mean value per year of the predicted number of FRs, from the number of CMEs in catalogs, (Table~\ref{Table:MCpredicted}, columns 2\todash 5) with the \insitu\ observed and corrected number of FRs (Table~\ref{Table:MCpredicted}, column 6).  A close matching is found between the numbers predicted from CDAW and derived from Lepping's list (370 and 390, respectively). A fair match is also found between the numbers
predicted from CACTus, SEEDs, and ARTEMIS (790, 740, and 720, respectively) and derived from Lynch's list (850).  The predicted yearly averaged numbers of FRs depend more on the type of catalog (manual/algorithm based) than on the period of time analyzed (Table~\ref{Table:MCpredicted}). Uncertainties on the number of CMEs/FRs, present in both CME catalogs and MCs lists, is due to thresholds implemented, which are more or less conservative depending on the authors' choice.  This leads to a factor $\approx$ two between the different estimations of the number of CMEs/MCs.  We also recall the use of \eq{dNsdRtot} with $\theta_{\rm max}=45 \degree$ to correct the observed number of FRs.  The quoted numbers would be simply multiplied by a factor 1.4 (resp. 0.8) if the extreme values $\theta_{\rm max}=30 \degree$ (resp. $60 \degree$) would be used instead, so this has a lower effect than the factor two quoted above. Indeed in the logarithm scale of \fig{dNsdRcumul}a, this corresponds to a shift of the curves of $0.15$ upward and $0.1$ downward, respectively, so a small shift compared to the variations of $\log_{10} N(R)$. 

Opposite to the case of MCs, the number of small FRs is much too large compared with the number of FRs predicted by the CME catalogs by a factor between 7 and 21 (Table~\ref{Table:MCpredicted}).  This conclusion is reinforced by the following two points: first, the fact that algorithm-based detections of CMEs have a fraction of false detections or split some events into several \cite{Yashiro08} implies that the quoted prediction of FRs in Table~\ref{Table:MCpredicted} with CACTus, SEEDS, and ARTEMIS are expected to be overestimated.  Second, there is no evidence that the power law found for small FRs would end or change to a lower slope at smaller radius (\sect{SFR}), then even more small FRs would be expected at smaller scales.   A way to decrease the number of small FRs would be to drastically reduced $\theta_{\rm max}$.   However, if these small FRs are of solar origin as argued in \sect{origin_SFR}), both blowout jets and streamers have a broad range of latitudes so $\theta_{\rm max}$ cannot be small.  The two above points imply that the predicted number of small FRs in Table~\ref{Table:MCpredicted} (column~7) from \inlinecite{Feng08} is expected to be a lower bound.

In summary, the number of CMEs found with an algorithm-based catalog is expected to be an overestimation, while the number of small FRs is expected to be an underestimation. 
Because the number of estimated FRs from CMEs is a factor $\approx$ ten times lower than the number of small FRs, and this factor is expected to be even higher as argued above, we conclude that the small FRs are not related to CMEs, even when including the narrow CMEs.   On the contrary, the predicted number of FRs from the CME catalogs match the corrected number of MCs within a factor two of uncertainty (depending on the strictness of the criteria to define both a CME and a MC).  We conclude that only the MCs, with a Gaussian-like distribution of radius, are the counter parts of all CMEs.  
 
\section{What Origins for the Small Flux Ropes?} 
      \label{sect:origin_SFR}      
      
Several scenarios for the formation of small flux ropes are possible, as follows.

First, lower in the solar atmosphere, blowout X-ray jets are jets with particular characteristics such as the ejective eruption of the magnetic arcade at their base \cite{Moore10}.  The destabilization of a closed-field configuration inferred in such processes is similar to that of the initiation of CMEs, the major difference being that reconnection takes place with an open magnetic field.  This so-called interchange reconnection removes parts of the overlying arcade, decreasing the downward magnetic tension, and then allowing the core to erupt in a jet-like manner.    MHD simulations reproducing the main characteristics of these blowout X-ray jets involve shearing motions of an arcade, or the emergence of a twisted flux tube, in an open field environment \cite{Pariat10,Archontis13}.   Evidence of the ejections of a flux rope is found both in these simulations and in observations \cite{Liu11,Moore13}. Then, while CMEs typically originate from the core of closed-field regions so that the erupting core should have enough magnetic flux to make its way through the overlying stabilizing field, a small magnetic bipole can have an ejective eruption if it is located in an open-field environment.  Since smaller magnetic bipoles are much more numerous than larger ones, blowout X-ray jets are natural candidates for the origin of small FRs in the interplanetary space.

Second, there is observational evidence of ejection of plasma blobs from above the cusp of streamers \cite{Sheeley97,Wang00,Sheeley09,Song09}. With STEREO imagers, these plasma blobs can be tracked into the inner heliosphere \cite{Sheeley10,Rouillard10}.
These transients have a large range of sizes and some have evidence of a twisted magnetic structure. They are therefore good candidates for the origin of small FRs.  3D MHD simulations, with settings tuned to the configuration above streamers, localize the main instability very close to the streamer cusp \cite{Chen09}.
Then, reconnection in the current sheet above streamers, with the presence of a guiding field along the current sheet, is also a candidate for the origin of small FRs in the interplanetary space. Indeed, MHD simulations show the formation of flux ropes from the development of the tearing mode \cite{Einaudi01}. More complex cases, involving several current sheets modeling adjacent helmet streamers, show the development of multiple tearing modes with growth rates much more important than the simple tearing-mode prediction \cite{Dahlburg95}. In their simulation, \inlinecite{Dahlburg95} showed that such a mechanism can lead to the formation of plasmoids with size scales between $3.5\times 10^{-5}$ AU and $3.5\times 10^{-3}$ AU when present at 1~AU. This upper limit agrees with the lower limit of very small flux ropes observed (\fig{dNsdR}). However, the origin of small flux ropes with larger sizes remains to be investigated.

A plausible alternative for the formation of small FRs further away from the Sun is the hypothesis that they originate from the heliospheric current sheet (HCS: \opencite{Moldwin00}). In such a scenario, the small FRs would result from multiple reconnections during the development of the tearing instability in the HCS. This process is similar to that advocated for the FR formation above streamers and for the ones observed in the Earth's magnetotail (\opencite{Linton09} and references therein).   However, similarly as for simulations for helmet streamers, the thickness of the HCS is too small to clearly explain the formation of structures as large as $0.05$~AU: the HCS is $\approx 10^4$~km at 1~AU, so less than $10^{-4}$~AU (\eg\ the review of \opencite{Smith01}, and more recently \opencite{Blanco06}). This is ten times lower than the smallest FRs detected.

Recently, efforts have been made in numerical simulations to investigate the properties of plasmoid chains, formed by the non-linear development of the tearing instability in current sheets. This development results in coalescence of plasmoids to form larger structures. A theoretical approach on the merging of those structures has linked the island half size [$w_{\rm isl}$] with the reconnection rate [$\epsilon$] and the system size $L$ as follows: $w_{\rm isl}=\epsilon L$ \cite{Fermo2010,Uzdensky2010}. This result has been numerically tested by \inlinecite{Loureiro2012} for high Lundquist-number plasmas, which is a necessary condition to form plasmoid chains in unstable current sheets (via the plasmoid instability; see \opencite{Loureiro2007}). In this work, the authors showed that the effective reconnection rate is of the order of $\epsilon \approx 0.02$ (independent of the Lundquist number for large values), leading to plasmoids of size of the order of $0.02\,L$. 
Then, in the context of HCS, a system size up to 1~AU can be considered, leading to flux ropes of radius 0.02 AU. Moreover, so-called ``monster'' plasmoids can form, of which the maximum size has been reported to be $\approx 0.1\, L$. Interestingly, \inlinecite{Loureiro2012} also found a plasmoid population characterized by a distribution function scaling as $w_{\rm isl}^{-2}$, which is comparable to the slope found in the present paper for the distribution of small flux ropes (\fig{dNsdRcor}).
It is however difficult to conclude with confidence that small flux ropes are of same nature as plasmoids formed by coalescence of smaller structures in current sheets, both in the HCS and in the coronal streamers. Indeed, simulation-based current sheets present a much simplier configuration compared to that in the heliosphere. The effect of the solar wind, for example, can stop the growth of FRs by pushing them away from their formation region. 


The most straightforward way to test the above three possibilities would be to follow \insitu\ detected small FRs with heliospheric imagers backward in time, and therefore toward the Sun.  The main
difficulty is that they are typically not seen in imagers because they are both small and not dense enough.  Even close to the Sun, the streamer blobs are only 10 \% over-dense compared to the surroundings. As they expand, while propagating away from the Sun, they have even less contrast, so that following these structures at $\approx 1$~AU becomes difficult. A possible way to overcome such difficulties would be to consider small FRs that have been swept by a high-speed stream, which makes them appear as the densest structures \cite{Rouillard11}. Over the six small FRs studied, only the four largest ones could be followed back to the Sun with the STEREO imagers.  Two have been identified as CMEs and the two others as streamers blobs.
Their half size at 1~AU is between 0.025 and 0.06~AU, so they are in the intermediate range 
where both small FR and MC distributions overlap (\fig{dNsdR}), which is consistent with their mixed solar origin.   The last two small FRs studied have a half size of $\approx$ 0.01 and 0.02~AU, but the source of these smaller FRs could not be identified with imagers.

Another possibility for the source of small FRs is provided by the analysis of the statistical distributions in size of these sources, and these can be compared to the results of \fig{dNsdRcor}. However, to the authors' knowledge, none of these distributions are known.  At best, one can use the distribution of CMEs that was derived by \inlinecite{Robbrecht09b} and that was extended  to small-scale eruptions by \inlinecite{Schrijver10}. 
A power law with a slope $\approx -1.8$ was derived for the apparent width of the events, which could be compared with \fig{dNsdRcor} (where the slope is $\approx -2.4$).  However, there is no bump in the CME part of the distribution, as would be expected from \fig{dNsdRcor} if MCs were closely associated to CMEs. Also, the two studies provide solar distributions on the maximum extension of the CMEs. This measurement is generally associated with the extension along the axis of a possible FR within the CME rather than the FR radius. Moreover, the maximum apparent width is affected by projection and selection effects which are different for coronagraph and EUV imagers. 
Then, the distribution of the angular width of the CMEs cannot be directly compared with the distribution of the radius of MCs (from \fig{dNsdRcor}), and it requires a much deeper analysis, including modeling, to have comparable physical quantities.

Power laws are found in several domains including flare energy, but this has no obvious link with the power law found for small FRs.   A plausibly related power law is the one found for the photospheric magnetic field (\opencite{Parnell09} and references therein), as follows.
The small bipoles on the Sun emerge and disappear at a rapid rate (with a life time typically less than a day).   For bipoles emerging in an open field, it is plausible that the interchange reconnection occurs a few times, or maybe only once, when the bipole has its maximum extension and before at least one of its photospheric polarity merges with the surrounding network field.   Then, if small FRs are formed by blowout X-ray jets, one can envision a close link between the distribution of the photospheric magnetic field and that of small FRs.   For CMEs that are dominantly coming from ARs, the link is less obvious.   First, there is an ongoing debate on the type of distributions for the photospheric magnetic field at the scale of ARs (\opencite{Zhang10} and references there in).  Second,  ARs, contrary to small bipoles, last for a longer time so that their magnetic configuration can be re-energized after a CME.  This implies a broad range of CMEs numbers launched from ARs, ranging from none to more than 60 \cite{Green02,Chen11}.  Therefore, the link between the distributions of the photospheric magnetic field and CMEs, and so MCs, is not direct, and the solar interpretation related to \fig{dNsdR_MCs} would need developments outside the scope of this article.

All in all, observational arguments and theoretical/numerical backgrounds are more likely to associate the origin of small FRs to blowout X-ray jets.  However, nonlinear developments of tearing modes in the cusp of streamers is a small FR formation scenario also supported by both observational and theoretical arguments.  Moreover, the number of small FRs peaks before crossing the sector boundary ($\approx$ six hours or less; see Figure~9 of \opencite{Cartwright10}), which is in agreement with a scenario for the formation of small FRs above streamers. Finally, there is presently no clear argument in favor of formation within the HCS.
   
\section{Conclusion} 
      \label{sect:Conclusion} 
 
 Following the controversy over whether two populations of FRs exist in the solar wind (\sect{Introduction}), we investigate the FR distributions from four published lists of events.
Detected flux ropes were all fitted by the same flux-rope model, \eq{Lundquist}, which provides an estimation of the FR parameters such as its radius. Due to the limited number of observed events in each list (between 121 and 144) and the 
huge variation of the number of cases along the range of the distributions, histograms classsically used for statistical study of distributions are too limited for an appropriate analysis here.   Then, we develop a partition method which has the main advantage of spreading uniformly the statistical fluctuations across the distribution .
This method was broadly tested with theoretical random distributions, in particular for distributions similar to the observed ones. We show that this new method is indeed robust even for the limited number of observed FRs presently available (\fig{dNsdRtest}). Investigating possible selection effects on the distributions of the flux-rope radius, we find biases in the observed distributions due to the orientation (\fig{lambda}) as well as the apparent surface of detected FRs. We propose a method applicable for any distribution of flux ropes to correct these biases. After these corrections, we derive the distribution of $\dNsdRtotSD$ versus $R$ where $\Ntot $ is the number of FRs per year at 1~AU and $R$ is the FR radius.
We find the following results:  
\begin{itemize}
 \item Two FRs populations with different distributions [$\dNsdRtotSD$]: a power law for small FRs and a Gaussian-like for MCs (\fig{dNsdRcor}).
  \item The power law for small FRs [$\dNsdRtotSD \propto R^{s}$] is a steep function of the FR radius [$R$] since the slope $s= -2.4 \pm 0.14$ for the period from 1995 to 2005. 
  \item There is evidence that the slope [$s$] is time dependent but the number of FRs and the time range are both too small to assert a link with the solar cycle.
  \item The MC distribution is Gaussian-like, closely centered around a FR radius of $0.1$~AU with a standard deviation slightly lower than $0.03$~AU (\fig{dNsdR_MCs}).
  \item The distributions of small FRs and MCs overlap around $R=0.05$ AU.  This overlap is limited in radius both because the distribution of small FRs is steeply decreasing function of $R$ and because it is also the case for the MC distribution for decreasing $R$ values (\fig{dNsdRcor}). 
\end{itemize}

Since they have remarkably different distributions, small FRs and MCs should have different origins.  On the one hand, we find that the number of FRs per year predicted from the observed CMEs is close to the observed number of MCs when the observed fraction of MCs in ICMEs is taken into account.   More precisely, the predicted number of MCs from the CDAW CME catalog is quite similar to the number of MCs per year present in the list of \inlinecite{Lepping10}, while two times more MCs are predicted from the algorithm-based CME catalogs (CACTus, SEEDS, and ARTEMIS), in agreement with the number of MCs per year present in the lists of \inlinecite{Lynch05} and \inlinecite{Feng07}. On the other hand, the small detected FRs are too numerous when compared to the observed CME rates, in particular presenting ratios by at least a factor of ten. These two comparisons of predicted MCs and small flux ropes lead to the conclusion that only the MCs, with a Gaussian-like distribution versus FR radius, are associated to observed CMEs. 
  

Since we have shown that the small FRs are not associated with CMEs, even narrow ones,  we propose possible scenarios for the formation of those small FRs.
These twisted structures may still come from other eruptive coronal phenomena such as blowout X-ray jets or multi-reconnection processes, likely to involve nonlinear tearing-mode-like phenomena, in the current sheet at the top of streamers.
 Both of these scenarios, explaining the solar origin of flux ropes, have some observational and theoretical results that favor either one of these processes to be at the origin of the small FRs observed at 1~AU.  Presently, a clear answer cannot be given since available instruments are unable to image these small FRs so as to trace back their propagation from the Sun. A possible approach, beyond the scope of this article, would require the derivation of the distribution of plasma-blob radius in the corona and comparing this distribution to that of the size of small FRs derived in this article.
Another possible scenario for the formation of small FRs is that they can be formed away from the Sun, in the heliospheric current sheet. Similarly to helmet streamers, such a scenario would involve nonlinear development of the tearing instability in multiple current sheets forming the HCS. Comparisons with simulations of the coalescence of plasmoid chains suggest a similar power law for the distribution of plasmoids as for the FR radius (\fig{dNsdRcor}b), as well as sizes of ``monster'' plasmoids that could explain the small FR investigated here. However, numerical simulations leave out more complex processes involving, for example, the solar wind, so that the formation of small FRs in the HCS remains unclear.
To support either of these scenarios, observations will need the capability to track and/or detect flux ropes closer to the Sun. The future Solar Orbiter spacecraft should be particularly well
suited for these purposes.


\begin{acks}
The present work was partially funded by a contract from the AXA Research Fund (MJ) and also supported by the Argentinean grant UBACyT 20020120100220 (SD) and by a one month invitation of SD by the Paris
Observatory.
SD is member of the Carrera del Investigador Científico, CONICET. 
SD acknowledges support from the Abdus Salam International Centre for Theoretical Physics (ICTP), as
provided in the framework of his regular associateship.
\end{acks}

   
\bibliographystyle{spr-mp-sola}

\bibliography{mc}  

\IfFileExists{\jobname.bbl}{} {\typeout{}
\typeout{****************************************************}
\typeout{****************************************************}
\typeout{** Please run "bibtex \jobname" to obtain} \typeout{**
the bibliography and then re-run LaTeX} \typeout{** twice to fix
the references !}
\typeout{****************************************************}
\typeout{****************************************************}
\typeout{}}

\end{article} 

\end{document}